# The transition from quantum field theory to one-particle quantum mechanics and a proposed interpretation of Aharonov-Bohm effect

**Benliang Li[1,2,*], Daniel W. Hewak[2], Qi Jie Wang[1,3]**

[1]*Centre for OptoElectronics and Biophotonics (COEB), School of Electrical & Electronic Engineering, The Photonics Institute, Nanyang Technological University, Singapore, 639798*

[2]*Optoelectronics Research Centre (ORC), University of Southampton, Southampton SO17 1BJ, United Kingdom*

[3]*Centre for Disruptive Photonic Technologies (CDPT), School of Physical and Mathematical Sciences, The Photonics Institute, Nanyang Technological University, Singapore, 637371*

[*]Email: libenliang732@gmail.com

In this article we demonstrate a sense in which the one-particle quantum mechanics (OPQM) and the classical electromagnetic four-potential arise from quantum field theory (QFT). In addition, the classical Maxwell equations are derived from a QFT scattering process, while both classical electromagnetic fields and potentials serve as mathematical tools to approximate the interactions among elementary particles described by QFT physics. Furthermore, a plausible interpretation of the Aharonov-Bohm (AB) effect is raised within the QFT framework. We provide a quantum treatment of the source of electromagnetic potentials and argue that the underlying mechanism in the AB effect can be understood via interactions among electrons described by QFT where the interactions are mediated by virtual photons.

A few remarks need to be recalled on the historical development of electromagnetics. In the 19th century, a description of electromagnetic phenomena was developed. According to that description, charges and currents act as local sources of force fields, and these force fields acted on other charges and currents locally through the Lorentz equation of motion. Meanwhile, the electromagnetic field was treated as a combined real physical entity to interpret the phenomena as local. In the 20th century, in order to describe the behaviour of the microscopic particles, physicists have developed the quantum field theory (QFT) which gives a full quantum treatment of the fields. They found that the fields are made of some elementary particles, i.e., Dirac field is made of electrons and positrons, and the electromagnetic field is made of photons. However, the classical electromagnetic fields in real situations can take various mathematical expressions; to date, the understanding of some classical fields within the framework of QFT is still unclear and challenging. In this article, we described the classical fields and potentials in terms of QFT mathematical languages.

The Aharonov-Bohm (AB) effect has attracted many researchers' interest due to its conceptual importance [1]. Since the electromagnetic potential is not gauge-invariant, it cannot represent a physical entity. Some researchers believe that the motion of a charged particle can be influenced by the electromagnetic fields confined to a region from which the particle is rigorously excluded, which raises some debates on the non-local feature in the quantum theory [2-10]. In this article, we focus more on revealing an alternative interpretation mechanism by considering the full quantum treatment on a source of the electromagnetic potential within the framework of quantum electrodynamics (QED), instead of joining the discussions over the non-local feature of one particle quantum mechanics (OPQM). Throughout the article, we abbreviated Eq. (A3) as EMF and Eq. (A8) as FEMF respectively, and also we used the natural units in which light velocity and Planck's constant equal to unity.

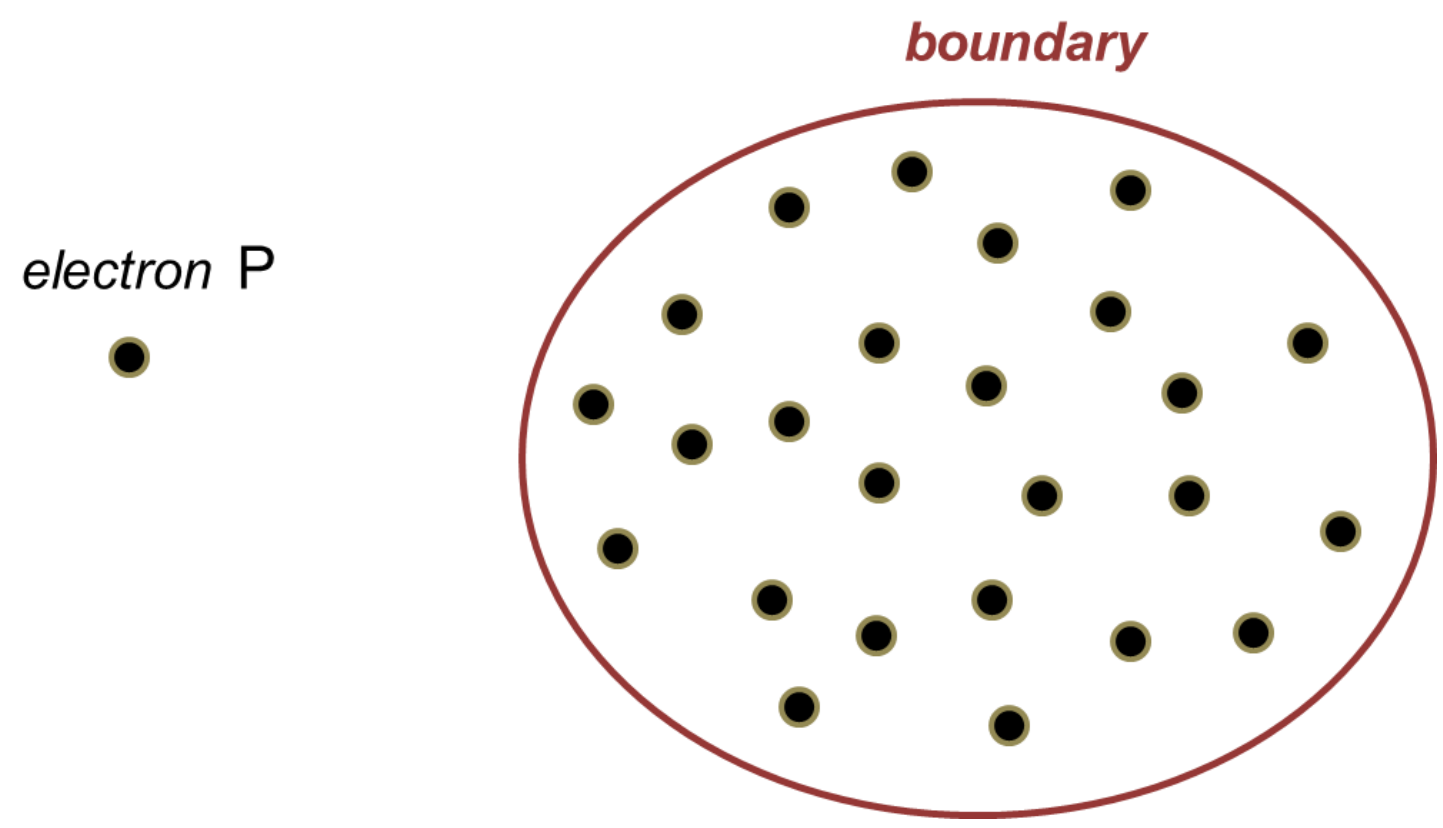

Fig. 1 Sketch of an ensemble of electrons which are confined within a boundary and the electron P which is outside of the boundary.

To get the solutions of the above problems, we are going to derive the classical four-potential from QED. The Hamiltonian of QED theory without FEMF is

$$\hat{H}_{QED} = \int d^3\boldsymbol{x}[\hat{\bar{\psi}}(-i\boldsymbol{\gamma}\cdot\vec{\nabla} - e\gamma^{\mu}\hat{A}_{\mu} + m_e)\hat{\psi}] \tag{1}$$

in which $\hat{\bar{\psi}} \equiv \hat{\psi}^{\dagger}\gamma^0$, $\hat{A}_{\mu} = (\hat{\phi}, -\hat{\boldsymbol{A}})$ is the quantum electromagnetic four-potential, $e$ is the coupling coefficient which takes a positive value, $m_e$ is electron's rest energy and $\gamma^{\mu}$ is a four Dirac Gamma matrices. In radiation gauge, the electromagnetic vector potential and Dirac electron field with fixed spin in the interaction picture are

$$\hat{\boldsymbol{A}}_I(\boldsymbol{x},t) = \int \frac{d^3\boldsymbol{k}}{(2\pi)^3}\frac{1}{\sqrt{2\omega_k}}\sum_{\lambda=1,2}[\hat{a}_{\boldsymbol{k}}^{\lambda}\boldsymbol{v}^{\lambda}e^{-i(\omega_k t - \boldsymbol{k}\cdot\boldsymbol{x})} + \hat{a}_{\boldsymbol{k}}^{\lambda\dagger}\boldsymbol{v}^{\lambda}e^{i(\omega_k t - \boldsymbol{k}\cdot\boldsymbol{x})}] \tag{2}$$

$$\hat{\psi}_I(\boldsymbol{x},t) = \int \frac{d^3\boldsymbol{k}}{(2\pi)^3}\frac{1}{\sqrt{2\omega_k}}\hat{c}_{\boldsymbol{k}}u_{\boldsymbol{k}}e^{-i(\omega_k t - \boldsymbol{k}\cdot\boldsymbol{x})} \tag{3}$$

where $\hat{a}_{\boldsymbol{k}}^{\lambda}$ and $\hat{c}_{\boldsymbol{k}}$ are photon and electron annihilation operators that satisfy commutation relation $[\hat{a}_{\boldsymbol{k}}^{\lambda}, \hat{a}_{\boldsymbol{k}'}^{\lambda'\dagger}] = (2\pi)^3 \delta^{\lambda\lambda'} \delta^3(\boldsymbol{k}-\boldsymbol{k}')$ and anti-commutation relation $\{\hat{c}_{\boldsymbol{k}}, \hat{c}_{\boldsymbol{k}'}^{\dagger}\} = (2\pi)^3 \delta^3(\boldsymbol{k}-\boldsymbol{k}')$, respectively. We also have $u_{\boldsymbol{k}} = \begin{pmatrix} \xi\sqrt{\omega_k - \boldsymbol{k}\cdot\boldsymbol{\sigma}} \\ \xi\sqrt{\omega_k + \boldsymbol{k}\cdot\boldsymbol{\sigma}} \end{pmatrix}$ where we take the positive root of each eigenvalue when taking the square root of the matrix in $u_{\boldsymbol{k}}$, and $\omega_{\boldsymbol{k}}^2 = m_e^2 + \boldsymbol{k}^2$ for on-shell electrons. The rest symbols are $\xi = \begin{pmatrix} 1 \\ 0 \end{pmatrix}$ and Pauli matrices $\boldsymbol{\sigma}$, $\boldsymbol{\nu}^{\lambda=1} = \frac{1}{\sqrt{k_1^2 + k_2^2}}(k_2, -k_1, 0)$ and $\boldsymbol{\nu}^{\lambda=2} = \frac{1}{\sqrt{(k_1^2 + k_2^2)|\boldsymbol{k}|^2}}(k_1 k_3, k_2 k_3, -k_1^2 - k_2^2)$ are two polarization directions of photons. To make the presentation simple, we omitted the positrons part in Eq. (3). The free and interaction Hamiltonians are $\hat{H}_0^{QED} = \int d^3\boldsymbol{x}[\hat{\bar{\psi}}(-i\boldsymbol{\gamma}\cdot\vec{\nabla} + m_e)\hat{\psi}]$ and $\hat{H}_{\text{int}}^{QED} = \int d^3\boldsymbol{x}(-e\hat{\bar{\psi}}\gamma^\mu \hat{A}_\mu \hat{\psi})$, respectively. In the interaction picture, the time evolution operator $\hat{U}_{QED}(t,t_0)$ obey equation $i\frac{\partial}{\partial t}\hat{U}_{QED}(t,t_0) = \hat{H}_I^{QED}(t)\hat{U}_{QED}(t,t_0)$ with $\hat{H}_I^{QED}(t) = e^{i\hat{H}_0^{QED}(t-t_0)}\hat{H}_{\text{int}}^{QED} e^{-i\hat{H}_0^{QED}(t-t_0)}$, and the time-dependent perturbation theory expanded to the second order gives

$$\hat{U}_{QED}(t,t_0) = \hat{S}_0^{QED}(t,t_0) + \hat{S}_1^{QED}(t,t_0) + \hat{S}_2^{QED}(t,t_0) + \cdots \tag{4}$$

in which $\hat{S}_1^{QED}(t,t_0) = (-i)\int_{t_0}^{t} dt_1 \hat{H}_I^{QED}(t_1)$ and $\hat{S}_2^{QED}(t,t_0) = (-i)^2\int_{t_0}^{t} dt_1 \int_{t_0}^{t_1} dt_2 \hat{H}_I^{QED}(t_1)\hat{H}_I^{QED}(t_2)$ are the first and second order perturbation terms, respectively. The zeroth order term $\hat{S}_0^{QED}(t,t_0) = 1$ is just an identity operator.

In OPQM framework, for one electron in the presence of a classical static four-potential $A_\mu(\boldsymbol{x})$, the OPQM Hamiltonian can be given as

$$H = \gamma^0[-i\boldsymbol{\gamma}\cdot\vec{\nabla} - e\gamma^\mu A_\mu(\boldsymbol{x}) + m_e] \tag{5}$$

in which the interaction energy is $H_{\text{int}}(\boldsymbol{x}) = -e\gamma^0\gamma^\mu A_\mu(\boldsymbol{x})$ with $A_\mu(\boldsymbol{x}) = [\phi(\boldsymbol{x}), -\boldsymbol{A}(\boldsymbol{x})]$ and the free energy is $H_0 = \gamma^0(-i\boldsymbol{\gamma}\cdot\vec{\nabla} + m_e)$. By comparison, we note that the photon field operator $\hat{A}_\mu$ in Eq. (1) has been replaced by a function $A_\mu(\boldsymbol{x})$ in Eq. (5); later on, we will explain how the function $A_\mu(\boldsymbol{x})$ arises from QED theory. In the interaction picture of OPQM, we have the interaction energy and the evolution operator as

$$
\begin{aligned}
&H_I(t) = \exp[iH_0(t-t_0)]H_{\text{int}}(\boldsymbol{x})\exp[-iH_0(t-t_0)] \\
&U(t,t_0) = T\left\{\exp[-i\int_{t_0}^{t} dt' H_I(t')]\right\}
\end{aligned}
\tag{6}
$$

in which $T$ stands for the time-ordering operator. Similar to Eq. (4), we can expand $U(t,t_0)$ in perturbation series up to the leading order which is the first order as

$$
U(t,t_0) = S_0(t,t_0) + S_1(t,t_0) + \cdots \tag{7}
$$

in which $S_0(t,t_0) = 1$ and $S_1(t,t_0) = (-i)\int_{t_0}^{t} dt_1 H_I(t_1)$. Note that we denote the unitary evolution operators as $\hat{U}_{QED}$ and $U$ for QED and OPQM theories, respectively. By integrating over an infinite time period, we can obtain the matrix elements of $U(t,t_0)$ expanded in momentum basis as $\langle \boldsymbol{p}' | U(\infty,-\infty) | \boldsymbol{p} \rangle$. After plugging $H_I(t)$ from Eq. (6) into $\langle \boldsymbol{p}' | U(\infty,-\infty) | \boldsymbol{p} \rangle$, we get results containing $2\pi\delta(\omega_{\boldsymbol{p}'} - \omega_{\boldsymbol{p}})$ which put an energy conservation constraint on the initial and final state of the electron P, this appears as a serious problem in the formulation of OPQM since we nearly get an identity matrix $U(\infty,-\infty)$. One may notice that this problem can be fixed by introducing another subsystem interacting with P, under this condition, the free energy $H_0$ in Eq. (6) will cover both the electron P and that subsystem, and the overall energy is conserved after the time integration. Bearing in mind that our target is to derive the classical four-potential $A_\mu(\boldsymbol{x})$ from QED process, therefore, we can just focus on the dynamical evolutions of the electron P described by QED and OPQM, respectively. We get the dynamical transition matrix expanded in momentum basis up to the first order of perturbations as

$$
\langle \boldsymbol{p}' | U(\infty,-\infty) | \boldsymbol{p} \rangle = S_0(\boldsymbol{p}',\boldsymbol{p}) + S_1(\boldsymbol{p}',\boldsymbol{p}) \tag{8}
$$

in which $S_1(\boldsymbol{p}',\boldsymbol{p}) \equiv -iC_1 \langle \boldsymbol{p}' | H_{\text{int}}(\boldsymbol{x}) | \boldsymbol{p} \rangle$ is the first order transition matrix. The constant factor $C_1$ carries an inverse energy dimension which may indicate the overall energy conservation obtained over the infinite-time-integration technique. For the zeroth order perturbation expansion $S_0(\boldsymbol{p}',\boldsymbol{p}) \equiv \langle \boldsymbol{p}' | \boldsymbol{p} \rangle$, if we attach the other subsystem such as $|\Psi\rangle$, we would obtain an additional term as $\langle \Psi | \Psi \rangle$. The information of a non-trivial dynamical transition between the states of electron P is carried by the first order perturbation term $S_1(\boldsymbol{p}',\boldsymbol{p})$.

As stated in Fig. 1, we name the single electron outside of the boundary as P. At time $t_0$, we denote the state of P with momentum $\boldsymbol{p}$ as $|\boldsymbol{p}\rangle \equiv \sqrt{2\omega_{\boldsymbol{p}}}c_{\boldsymbol{p}}^{\dagger}|0\rangle$ in which $|0\rangle$ is the vacuum state, $c_{\boldsymbol{p}}^{\dagger}$ is the creation operator of electrons and $\omega_{\boldsymbol{p}}^2 = m_e^2 + \boldsymbol{p}^2$ is the energy of electron P. Moreover, at time $t_0$, the ensemble of electrons confined within a boundary shown in Fig. 1 can be given by state $|\Psi\rangle$. Therefore, the whole system that includes P and the ensemble of electrons can be expressed as a quantum state $|\Psi, \boldsymbol{p}\rangle$. We assume that the interaction energy between P and $|\Psi\rangle$ is much smaller than the free energy of P (this is due to the macroscopic distance between P and $|\Psi\rangle$), then the evolution of P can be known theoretically, using perturbation theory. We further assume that the combined system $|\Psi, \boldsymbol{p}\rangle$ is kept in an isolated situation and the ensemble of the electrons is in a macroscopically equilibrium state; therefore we expect that the state $|\Psi\rangle$ does not vary macroscopically throughout the whole experimental time. Such physical idealization is a good approximation provided that, there are no dramatic disturbances caused by the environment and no significant external forces acted on $|\Psi, \boldsymbol{p}\rangle$ from other systems. Later on we will see that this approximation can lead us to derive a non-time-varying classical potential $A_\mu(\boldsymbol{x})$. In order to make a comparison of the dynamical evolution of the electron P between QED and OPQM formulations, we need to apply the same method to remove the time-dependency in Eq. (4). From the interaction picture, we observed that the field operators of Eq. (2) and Eq. (3) display a rotating-time-dependency through the mathematical expression $e^{i\omega t}$, for the macroscopically stationary system $|\Psi\rangle$, this rotating-time-dependency would allow us to make an approximation as $\hat{U}_{QED}(t+\mathrm{T}, t_0)|\Psi, \boldsymbol{p}\rangle = \hat{U}_{QED}(t, t_0)|\Psi, \boldsymbol{p}\rangle$ in which $\mathrm{T} \gg 2\pi/\omega$, this means that the effects on P caused by the variations of any electrons' state inside of $|\Psi\rangle$ would be cancelled away in a long-term by other variations within the system $|\Psi\rangle$ itself; this is the mean field approximation. Therefore we can remove this rotating-time-dependence by integrating over an infinite time period; this integration would usually generate an overall energy conservation constraint given by a Dirac delta function.

As discussed above, there are two methods we can use to describe the evolution of electron P shown in Fig. 1. The first one is based on the QED theory. According to QED, each electron inside of the boundary will interact with electron P where the interaction is mediated by virtual photons. Once we compute all the Feynman diagrams, the evolution of electron P will be known. The second method is based on OPQM. According to classical physics, each electron inside of the boundary will create an electromagnetic four-potential $A^{\mu}$ at the location of electron P. Once we sum over the four-potential created by every electron inside of the boundary and plug the final expression $A^{\mu}(\boldsymbol{x})$ into Eq. (5), the evolution of electron P will be known by computing the OPQM evolution operator $U$ . Indeed, these two methods must yield the same result since they are both correct. Thus, we can write

$$\hat{U}_{QED}(t,t_0)\left|\Psi,\boldsymbol{p}\right\rangle=\hat{U}_{QED}(t,t_0)\left|\Psi\right\rangle\otimes U(t,t_0)\left|\boldsymbol{p}\right\rangle \tag{9}$$

in which $U(t,t_0)$ is Eq. (7) and $\hat{U}_{QED}(t,t_0)$ is Eq. (4), $\left|\Psi,\boldsymbol{p}\right\rangle$ is the state of the whole system at initial time $t_0$ . Note that the operator $\hat{U}_{QED}(t,t_0)$ on the left-hand side of Eq. (9) will act on the whole system $\left|\Psi,\boldsymbol{p}\right\rangle$ while the operator $\hat{U}_{QED}(t,t_0)$ on the right-hand side of Eq. (9) only act on $\left|\Psi\right\rangle$. The right hand side of Eq. (9) is a tensor product of two systems which are $\left|\Psi\right\rangle$ and the electron P. The left-hand side of Eq. (9) is the evolution of the combined system $\left|\Psi,\boldsymbol{p}\right\rangle$ from the initial time $t_0$ to any later time $t$ governed by the QED theory. On the right hand side of Eq. (9), the evolution of $\left|\Psi\right\rangle$ and the evolution of P are governed by QED theory and OPQM theory respectively. These two kinds of evolution descriptions of the total system $\left|\Psi,\boldsymbol{p}\right\rangle$ from initial time $t_0$ to any later time $t$ need to be equivalent. That is, the final state of the total system $\left|\Psi,\boldsymbol{p}\right\rangle$ at any time $t$ given by the two different theories needs to be the same up to a global phase factor which is not measurable in quantum theory (here we just set the global phase difference to be zero). Practically, if we are interested only in the evolution of P, then we need to construct the OPQM theory to describe P which is $U\left|\boldsymbol{p}\right\rangle$. However, we still get another system $\left|\Psi\right\rangle$, and we just do tensor product it with P. The Eq. (9) holds good under the condition that the electron P is distinguishable from any electron inside the system $\left|\Psi\right\rangle$. This means that the state of P is orthogonal with any electron's state inside of $\left|\Psi\right\rangle$ during the evolution and, the exchanging between the electron P with any electron inside of $\left|\Psi\right\rangle$ is not allowed. Such separable condition is satisfied since the electron P was not in contact with the system $\left|\Psi\right\rangle$ during the evolution as depicted in Fig. 1. Moreover, the Eq. (9) holds good under the approximation that the influence on system $|\Psi\rangle$ acted by the electron P is negligible. Therefore, the evolution of the system $|\Psi\rangle$ is fully controlled by the internal interactions expressed by $\hat{U}_{QED}|\Psi\rangle$. Indeed, for a macroscopic system $|\Psi\rangle$ involving $N$ electrons, the evolution of any electron inside of $|\Psi\rangle$ is controlled by the other $N-1$ electrons plus the electron P, therefore, it would be reasonable that we only omit the effect caused by one single electron P in case that $N\to\infty$. As a matter of fact, for a charged particle moving in a classical potential created by other charged particles, the back action in OPQM framework is already neglected if the classical potential is given as Coulomb potential. Therefore, in order to reformulate the OPQM from QED, the back action needs to be neglected correspondingly. The Eq. (9) will not be valid in case of time-varying states $|\Psi(t)\rangle$ driven by some other external forces generated by a third party system interacting with $|\Psi\rangle$. This is due to the fact that the evolution of system $|\Psi\rangle$ cannot be expressed as $\hat{U}_{QED}|\Psi\rangle$ in the presence of some external forces. Meanwhile, the influence on P caused by the system $|\Psi\rangle$ is expressed by the OPQM evolution operator $U$, therefore, this $U$ is dependent on the system $|\Psi\rangle$ and our target is to construct the appropriate $U$, to satisfy the Eq. (9). We now product the state $\langle \boldsymbol{p}',\Psi|\equiv\langle 0|\sqrt{2\omega_{p'}}c_{p'}\otimes\langle\Psi|$ from left on both sides of Eq. (9) and perform the infinite-time-integration on both sides to get

$$\langle \boldsymbol{p}',\Psi|\hat{U}_{QED}(\infty,-\infty)|\Psi,\boldsymbol{p}\rangle=\langle\Psi|\hat{U}_{QED}(\infty,-\infty)|\Psi\rangle\langle\boldsymbol{p}'|U(\infty,-\infty)|\boldsymbol{p}\rangle \tag{10}$$

in which $\langle \boldsymbol{p}',\Psi|$ is the tensor product of a state of P with momentum $\boldsymbol{p}'$ and the initial state of system $|\Psi\rangle$, the right-hand side can be given as the product of two matrix elements under the separable condition. The Eq. (10) means that the two probability amplitudes, expressed by the two methods, of the transition between the initial state $|\Psi,\boldsymbol{p}\rangle$ to a final state $|\Psi,\boldsymbol{p}'\rangle$ need to be equal. As a simple example, the electrons state $|\Psi\rangle$ can be given as

$$|\Psi\rangle=2^{\frac{N}{2}}\sqrt{\omega_{k_1}\omega_{k_2}\cdots\omega_{k_N}}\,c_{k_1}^{\dagger}c_{k_2}^{\dagger}\cdots c_{k_N}^{\dagger}|0\rangle \tag{11}$$

in which $c_{k_j}^{\dagger}$ ( $j=1,2,\cdots,N$ ) represents the creation operator of a free electron state with momentum $\boldsymbol{k}_j$. Here we have a condition that any two electrons' states are orthogonal, i.e., $\langle 0 | c_{\boldsymbol{k}_i} c^\dagger_{\boldsymbol{k}_j} | 0 \rangle \propto \delta_{ij}$, by Pauli exclusion principle. From Eq. (4), the left-hand side of Eq. (10) can be written as

$$\langle \boldsymbol{p}', \Psi | \hat{U}_{QED}(\infty, -\infty) | \Psi, \boldsymbol{p} \rangle = \langle \boldsymbol{p}', \Psi | \hat{S}^{QED}_0 | \Psi, \boldsymbol{p} \rangle + \langle \boldsymbol{p}', \Psi | \hat{S}^{QED}_2(\infty, -\infty) | \Psi, \boldsymbol{p} \rangle \tag{12}$$

in which $\langle \boldsymbol{p}', \Psi | \hat{S}^{QED}_0 | \Psi, \boldsymbol{p} \rangle = \langle \Psi | \Psi \rangle \langle \boldsymbol{p}' | \boldsymbol{p} \rangle$ under the separable condition, note that the first order $\langle \boldsymbol{p}', \Psi | \hat{S}^{QED}_1(\infty, -\infty) | \Psi, \boldsymbol{p} \rangle = 0$. The second order term can be written as two parts, that is,

$$\begin{aligned} &\langle \boldsymbol{p}', \Psi | \hat{S}^{QED}_2(\infty, -\infty) | \Psi, \boldsymbol{p} \rangle \\ &= \langle \Psi | \hat{S}^{QED}_2(\infty, -\infty) | \Psi \rangle \langle \boldsymbol{p}' | \boldsymbol{p} \rangle + \sum_{i=1}^{N} \sum_{j=1}^{N} (-1)^{i+j} \langle \boldsymbol{p}', \boldsymbol{k}_i | \hat{S}^{QED}_2(\infty, -\infty) | \boldsymbol{k}_j, \boldsymbol{p} \rangle \langle \Psi - i | \Psi - j \rangle \end{aligned} \tag{13}$$

in which $| \boldsymbol{k}_j, \boldsymbol{p} \rangle \equiv 2\sqrt{\omega_{\boldsymbol{p}} \omega_{\boldsymbol{k}_j}} c^\dagger_{\boldsymbol{k}_j} c^\dagger_{\boldsymbol{p}} | 0 \rangle$ and $\langle \boldsymbol{p}', \boldsymbol{k}_i | \equiv \langle 0 | 2\sqrt{\omega_{\boldsymbol{p}'} \omega_{\boldsymbol{k}_i}} c_{\boldsymbol{p}'} c_{\boldsymbol{k}_i}$. The first term on the right-hand side of Eq. (13) is the internal interactions inside of system $| \Psi \rangle$ leaving the electron P unaffected, and the second term express the interactions between P and the system $| \Psi \rangle$, later on, we will see that the first order OPQM transition matrix arises from this term. $| \Psi - j \rangle$ is defined as the state of the remaining system after the removal of $| \boldsymbol{k}_j \rangle$, that is,

$$| \Psi - j \rangle \equiv 2^{\frac{N-1}{2}} \sqrt{\omega_{\boldsymbol{k}_1} \ldots \omega_{\boldsymbol{k}_{j-1}} \omega_{\boldsymbol{k}_{j+1}} \cdots \omega_{\boldsymbol{k}_N}} c^\dagger_{\boldsymbol{k}_1} \cdots c^\dagger_{\boldsymbol{k}_{j-1}} c^\dagger_{\boldsymbol{k}_{j+1}} \cdots c^\dagger_{\boldsymbol{k}_N} | 0 \rangle \tag{14}$$

Similarly, the state $| \Psi - i \rangle$ is

$$\langle \Psi - i | \equiv \langle 0 | 2^{\frac{N-1}{2}} \sqrt{\omega_{\boldsymbol{k}_1} \ldots \omega_{\boldsymbol{k}_{i-1}} \omega_{\boldsymbol{k}_{i+1}} \cdots \omega_{\boldsymbol{k}_N}} c_{\boldsymbol{k}_N} \cdots c_{\boldsymbol{k}_{i+1}} c_{\boldsymbol{k}_{i-1}} \cdots c_{\boldsymbol{k}_1} \tag{15}$$

The right-hand side of Eq. (10) expanded in leading order of perturbations can be given as

$$\begin{aligned} &\langle \boldsymbol{p}' | U(\infty, -\infty) | \boldsymbol{p} \rangle \langle \Psi | \hat{U}_{QED}(\infty, -\infty) | \Psi \rangle \\ &= \langle \boldsymbol{p}' | \boldsymbol{p} \rangle \langle \Psi | \Psi \rangle + \langle \boldsymbol{p}' | \boldsymbol{p} \rangle \langle \Psi | \hat{S}^{QED}_2(\infty, -\infty) | \Psi \rangle + \langle \boldsymbol{p}' | S_1(\infty, -\infty) | \boldsymbol{p} \rangle \langle \Psi | \Psi \rangle \end{aligned} \tag{16}$$

Compare Eq. (16) with Eq. (12) and Eq. (13), we get

$$\langle \boldsymbol{p}' | S_1(\infty, -\infty) | \boldsymbol{p} \rangle = \sum_{i=1}^{N} \sum_{j=1}^{N} (-1)^{i+j} \frac{\langle \Psi - i | \Psi - j \rangle}{\langle \Psi | \Psi \rangle} \langle \boldsymbol{p}', \boldsymbol{k}_i | \hat{S}^{QED}_2(\infty, -\infty) | \boldsymbol{k}_j, \boldsymbol{p} \rangle \tag{17}$$

For the terms with $i \neq j$, $\langle \Psi - i | \Psi - j \rangle = 0$ from the Pauli exclusion principle, therefore, we get

$$\langle \boldsymbol{p}' | S_1(\infty, -\infty) | \boldsymbol{p} \rangle = \sum_{j=1}^{N} \frac{\langle \boldsymbol{p}', \boldsymbol{k}_j | \hat{S}^{QED}_2(\infty, -\infty) | \boldsymbol{k}_j, \boldsymbol{p} \rangle}{\langle \boldsymbol{k}_j | \boldsymbol{k}_j \rangle} \tag{18}$$

At the right-hand side of Eq. (18), the incoming and outgoing states are $\left|\boldsymbol{k}_j,\boldsymbol{p}\right\rangle = 2\sqrt{\omega_{\boldsymbol{p}}\omega_{\boldsymbol{k}_j}}\,c_{\boldsymbol{p}}^{\dagger}c_{\boldsymbol{k}_j}^{\dagger}\left|0\right\rangle$ and $\left\langle \boldsymbol{p}',\boldsymbol{k}_j\right| = \left\langle 0\right|2\sqrt{\omega_{\boldsymbol{p}'}\omega_{\boldsymbol{k}_j}}\,c_{\boldsymbol{k}_j}c_{\boldsymbol{p}'}$, respectively. Note that since we neglected the back-action on $\left|\Psi\right\rangle$ caused by electron P as discussed in Eq. (9), as a consequence, we also need to overlook the energy-momentum conservation constraint on the term $\left\langle \boldsymbol{p}',\boldsymbol{k}_j\right|\hat{S}_2^{QED}(\infty,-\infty)\left|\boldsymbol{k}_j,\boldsymbol{p}\right\rangle$. The Eq. (18) indeed displays a very clear physical picture, the potential energy of P in the presence of $N$ electrons is the summation of individual contributions of the electrons. We can also find its counterpart in classical physics, i.e., the electrostatic potential energy of a point charge $q$ in the presence of other $N$ point charges equals to the summation over all contributions of $N$ point charges. The classical charged particles are well localized in space, this localization condition can be satisfied by demanding that the state $\left|\boldsymbol{k}_j\right\rangle$ of any electron is confined in a different small volume $x_j^3$. Therefore, we perform the infinite-time-integration on the right-hand side of Eq. (18) and obtain

$$\left\langle \boldsymbol{p}'\right|S_1(\infty,-\infty)\left|\boldsymbol{p}\right\rangle = \sum_{j=1}^{N}\frac{-i(2\pi)^4 e^2\exp[-i(\boldsymbol{p}'-\boldsymbol{p})\cdot\boldsymbol{x}_j]}{\left\langle \boldsymbol{k}_j\middle|\boldsymbol{k}_j\right\rangle[(\omega_{\boldsymbol{p}}-\omega_{\boldsymbol{p}'})^2-\left|\boldsymbol{p}-\boldsymbol{p}'\right|^2]}(\bar{u}_{\boldsymbol{p}'}\gamma^0 u_{\boldsymbol{p}}\bar{u}_{\boldsymbol{k}_j}\gamma^0 u_{\boldsymbol{k}_j}-\bar{u}_{\boldsymbol{p}'}\gamma^i u_{\boldsymbol{p}}\bar{u}_{\boldsymbol{k}_j}\gamma^i u_{\boldsymbol{k}_j}) \quad (19)$$

in which the factor $\exp[-i(\boldsymbol{p}'-\boldsymbol{p})\cdot\boldsymbol{x}_j]$ is added to indicate that each electron's state $\left|\boldsymbol{k}_j\right\rangle$ in the ensemble is well localized in a small volume $x_j^3$ and we also have $\left\langle \boldsymbol{k}_j\middle|\boldsymbol{k}_j\right\rangle = (2\pi)^3 2\omega_{\boldsymbol{k}_j}$. Thus, in order to determine the mathematical structure of $A_\mu(\boldsymbol{x})$, we can divide $\left\langle \boldsymbol{p}'\right|S_1(\infty,-\infty)\left|\boldsymbol{p}\right\rangle$ into two terms as

$$\left\langle \boldsymbol{p}'\right|S_1(\infty,-\infty)\left|\boldsymbol{p}\right\rangle = ieC_1(\left\langle \boldsymbol{p}'\right|\phi(\boldsymbol{x})\left|\boldsymbol{p}\right\rangle - \left\langle \boldsymbol{p}'\right|\gamma^0\boldsymbol{\gamma}\cdot\boldsymbol{A}(\boldsymbol{x})\left|\boldsymbol{p}\right\rangle) \quad (20)$$

Note that in this case $C_1 = 2\pi\delta(\omega_{\boldsymbol{p}'}+\omega_{\boldsymbol{k}'}-\omega_{\boldsymbol{p}}-\omega_{\boldsymbol{k}})$ and the state of the electron P is $\left\langle \boldsymbol{x}\middle|\boldsymbol{p}\right\rangle = u_{\boldsymbol{p}}e^{i\boldsymbol{p}\cdot\boldsymbol{x}}$ with $\left\langle \boldsymbol{x}\right| \equiv \left\langle 0\right|\hat{\psi}(\boldsymbol{x})$ and $\left|\boldsymbol{p}\right\rangle \equiv \sqrt{2\omega_{\boldsymbol{p}}}\,c_{\boldsymbol{p}}^{\dagger}\left|0\right\rangle$. Hence, apply the Feynman gauge and the non-relativistic limit $\omega_{\boldsymbol{p}}-\omega_{\boldsymbol{p}'}\approx 0$ in Eq. (19), after a short algebra, we can obtain the classical multi-particle potentials by comparing Eq. (19) with Eq. (20) as

$$\begin{aligned}\phi(\boldsymbol{x}) &= \sum_{j=1}^{N}\frac{-e\bar{u}_{\boldsymbol{k}_j}\gamma^0 u_{\boldsymbol{k}_j}}{2\omega_{\boldsymbol{k}_j}}\int\frac{d^3\boldsymbol{q}}{(2\pi)^3}\frac{1}{\left|\boldsymbol{q}\right|^2}\exp[i\boldsymbol{q}\cdot(\boldsymbol{x}-\boldsymbol{x}_j)] \\ A^i(\boldsymbol{x}) &= \sum_{j=1}^{N}\frac{-e\bar{u}_{\boldsymbol{k}_j}\gamma^i u_{\boldsymbol{k}_j}}{2\omega_{\boldsymbol{k}_j}}\int\frac{d^3\boldsymbol{q}}{(2\pi)^3}\frac{1}{\left|\boldsymbol{q}\right|^2}\exp[i\boldsymbol{q}\cdot(\boldsymbol{x}-\boldsymbol{x}_j)]\end{aligned} \quad (21)$$

Note that from Eq. (18) to Eq. (19), we excluded the exchange interaction. That is, the transition from $\boldsymbol{p}\to\boldsymbol{k}'$ and $\boldsymbol{k}\to\boldsymbol{p}'$, since this transition is not allowed due to the constraint that we impose on the system $\left|\Psi,\boldsymbol{p}\right\rangle$ which is that P can be distinguished from any electron in $\left|\Psi\right\rangle$, i.e., they are separated by a boundary. Note that the classical four-potential given above is negative due to the positive coupling constant $e$, this agrees with what we have been taught in classical physics: the electrons which carry negative charges create negative potentials. In the above expression of $A^{\mu}(\boldsymbol{x})$, we also notice that the term $(\omega_{\boldsymbol{p}}-\omega_{\boldsymbol{p}'})^2$ from the denominator of Eq. (19) is neglected, this is due to the fact that the potential $A^{\mu}(\boldsymbol{x})$ cannot be formulated as space-coordinate functions by including $(\omega_{\boldsymbol{p}}-\omega_{\boldsymbol{p}'})^2$. Therefore, the mathematical structure of OPQM fails to provide a precise description of the high energy particles' interactions since the potential $A^{\mu}(\boldsymbol{x})$ cannot be well defined in the high energy domain and it only arise from low energy physical phenomena, more discussions over such point of view can be seen in Appendix A. With the application of the Fourier expansion formula $\frac{1}{4\pi|\boldsymbol{x}|}=\int\frac{d^3\boldsymbol{p}}{(2\pi)^3}\frac{\exp(i\boldsymbol{p}\cdot\boldsymbol{x})}{|\boldsymbol{p}|^2}$, Eq. (21) can be given by a more elegant form as

$$A^{\mu}(\boldsymbol{x})=\sum_{j=1}^{N}\frac{-e\bar{u}_{\boldsymbol{k}_j}\gamma^{\mu}u_{\boldsymbol{k}_j}}{8\pi\omega_{\boldsymbol{k}_j}\left|\boldsymbol{x}-\boldsymbol{x}_j\right|} \tag{22}$$

And the electron static four-current can be given as

$$j^{\mu}(\boldsymbol{x})\equiv\sum_{j=1}^{N}\frac{-e}{2\omega_{\boldsymbol{k}_j}}\int\frac{d^3\boldsymbol{q}}{(2\pi)^3}\bar{u}_{\boldsymbol{k}_j}(\gamma^{\mu}-\frac{\boldsymbol{\gamma}\cdot\boldsymbol{q}}{\boldsymbol{q}})u_{\boldsymbol{k}_j}\exp[i\boldsymbol{q}\cdot(\boldsymbol{x}-\boldsymbol{x}_j)] \tag{23}$$

which gives the static classical Maxwell equations as

$$\vec{\nabla}^2A^{\mu}(\boldsymbol{x})-\vec{\nabla}\vec{\nabla}\cdot\boldsymbol{A}(\boldsymbol{x})=-j^{\mu}(\boldsymbol{x}) \tag{24}$$

Note that, we have the expression $\gamma^{\mu}-\frac{\boldsymbol{\gamma}\cdot\boldsymbol{q}}{\boldsymbol{q}}$ inside the four-current $j^{\mu}_{\boldsymbol{k}}(\boldsymbol{x})$ given above, that is, we subtract the component which is parallel to the momentum $\boldsymbol{q}$ to ensure the total classical charge conservation given as $\partial_{\mu}j^{\mu}_{\boldsymbol{k}}(\boldsymbol{x})=0$.

As we can see, the expression $A^{\mu}(\boldsymbol{x})$ in Eq. (22) carries a state $\left|\boldsymbol{k}_j\right\rangle$ dependency, therefore, the structure of classical four-potential $A^{\mu}(\boldsymbol{x})$ depends on the configuration of $\left|\Psi\right\rangle$. The "source" of the stationary classical electromagnetic four-potential $A_{\mu}(\boldsymbol{x})$ is found to be a relativistic quantum field, i.e., the Dirac electron field which we denoted as state $\left|\Psi\right\rangle$ in this scenario, the mathematical expressions of $A_{\mu}(\boldsymbol{x})$ appear to be completely arbitrary in

real situations due to the fact that the limitless configurations of the state $|\Psi\rangle$ can be found in classical world. Note that the infinite-time-integration approach cannot be applied in case of a time-varying function $A_\mu(t,\boldsymbol{x})$ to the derivations. Indeed, we expect to obtain a time-varying function $A_\mu(t,\boldsymbol{x})$ in case of a macroscopically varying state $|\Psi(t)\rangle$ driven by some external forces and such external forces would be generated by a third party system interacting with $|\Psi\rangle$, some specific examples with external forces can be a subject for future investigations.

Once we obtain the mathematical form of $A_\mu(t,\boldsymbol{x})$, then the EMF can be introduced by the following relations

$$\begin{aligned} \boldsymbol{B}(t,\boldsymbol{x}) &\equiv \vec{\nabla}\times\boldsymbol{A}(t,\boldsymbol{x}) \\ \boldsymbol{E}(t,\boldsymbol{x}) &\equiv -\vec{\nabla}\phi(t,\boldsymbol{x}) - \frac{\partial \boldsymbol{A}(t,\boldsymbol{x})}{\partial t} \end{aligned} \tag{25}$$

We note that these defined quantities $\boldsymbol{B}(t,\boldsymbol{x})$ and $\boldsymbol{E}(t,\boldsymbol{x})$ do not play any roles during our reformulation of OPQM from QED. The classical four-potential $A_\mu(t,\boldsymbol{x})$, as well as EMF defined above, are nothing but mathematical idealizations that approximate the interactions among electrons which are mediated by the virtual photons. The EMF defined above will play a role in classical theory since they comprise major parts of the Lorentz force equation. The classical fields $\boldsymbol{E}(\boldsymbol{x})$ and $\boldsymbol{B}(\boldsymbol{x})$ can be introduced from Eq. (22) following the relations defined in Eq. (25) as

$$\begin{aligned} \boldsymbol{E}(\boldsymbol{x}) &= \sum_{j=1}^{N} \frac{-e(\boldsymbol{x}-\boldsymbol{x}_j)}{4\pi\left|\boldsymbol{x}-\boldsymbol{x}_j\right|^3} \\ \boldsymbol{B}(\boldsymbol{x}) &= \sum_{j=1}^{N} \frac{-e\bar{u}_{\boldsymbol{k}_j}\boldsymbol{\gamma}\times(\boldsymbol{x}-\boldsymbol{x}_j)u_{\boldsymbol{k}_j}}{8\pi\omega_{\boldsymbol{k}_j}\left|\boldsymbol{x}-\boldsymbol{x}_j\right|^3} \end{aligned} \tag{26}$$

Note that the electric field does not depend on the energy of the electrons in the ensemble due to $\bar{u}_{\boldsymbol{k}_j}\gamma^0 u_{\boldsymbol{k}_j} = 2\omega_{\boldsymbol{k}_j}$. Again, the classical fields given in Eq. (26) will play significant roles in classical physics formulation as Lorentz equation of motion and their role is insignificant in the theoretical formulation of both QED and OPQM. We can further obtain the static Maxwell equations written in EMF form as

$$\begin{aligned} \vec{\nabla}\cdot\boldsymbol{E}(\boldsymbol{x}) &= j^0(\boldsymbol{x}) \\ \vec{\nabla}\times\boldsymbol{B}(\boldsymbol{x}) &= \vec{j}(\boldsymbol{x}) \end{aligned} \tag{27}$$

which are Gauss' Law and Ampere's Law respectively for static classical electromagnetic fields.

Coming to the analysis of QED theory on AB effect, in the schematic of double-slit experiment in which the AB effect can be observed as depicted in Fig. 2, the EMF is confined in the cylindrical solenoid. Suppose that this EMF is originally from spin and motion effects of an ensemble of electrons which are confined in the solenoid. In principle, this enormous number of electrons can be mathematically constructed as a quantum state $|\Psi\rangle$. And we denote the initial state of an electron with momentum $\boldsymbol{p}$ traveling outside of the solenoid as $|\boldsymbol{p}\rangle \equiv \sqrt{2\omega_{\boldsymbol{p}}} c_{\boldsymbol{p}}^{\dagger} |0\rangle$ in which $|0\rangle$ is the vacuum state. Here, we also named this traveling electron as P in all subsequent discussions. Note that for AB effect, we have the condition that the quantum state $|\Psi\rangle$ is confined in the solenoid and the electron P does not penetrate into the solenoid; this is the separable constraint that we mentioned in Fig. 1. That is, the electron P is distinguishable from each electron in the system $|\Psi\rangle$. It is clear that the constructing of an exact mathematical expression of $|\Psi\rangle$, which involves a macroscopic ensemble of electrons in a real world, would be a highly non-trivial task. However, our target here is to provide a qualitative analysis of what happens for AB effect using QED theory. We can obtain the evolution of system $|\Psi, \boldsymbol{p}\rangle$ up to the second order of perturbations as

$$\begin{aligned} &\langle \boldsymbol{p}', \Psi' | \hat{U}_{QED}(t, t_0) | \Psi, \boldsymbol{p} \rangle = \langle \boldsymbol{p}', \Psi' | \Psi, \boldsymbol{p} \rangle + \\ &(-i)\int_{t_0}^{t} dt_1 \langle \boldsymbol{p}', \Psi' | \hat{H}_I^{QED}(t_1) | \Psi, \boldsymbol{p} \rangle + (-i)^2 \int_{t_0}^{t} dt_1 \int_{t_0}^{t_1} dt_2 \langle \boldsymbol{p}', \Psi' | \hat{H}_I^{QED}(t_1) \hat{H}_I^{QED}(t_2) | \Psi, \boldsymbol{p} \rangle \end{aligned} \tag{28}$$

Since we do not include the photon field into the system, the first order perturbation term $\langle \boldsymbol{p}', \Psi' | \hat{H}_I^{QED}(t_1) | \Psi, \boldsymbol{p} \rangle = 0$. In the second order expression, we have

$$\begin{aligned} &\langle \boldsymbol{p}', \Psi' | \hat{H}_I^{QED}(t_1) \hat{H}_I^{QED}(t_2) | \Psi, \boldsymbol{p} \rangle = \\ &\langle \boldsymbol{p}', \Psi' | \int d^3\boldsymbol{x} d^3\boldsymbol{y} [e^2 \hat{\bar{\psi}}_I(\boldsymbol{x}, t_1) \gamma_\mu \hat{A}_I^\mu(\boldsymbol{x}, t_1) \hat{\psi}_I(\boldsymbol{x}, t_1) \hat{\bar{\psi}}_I(\boldsymbol{y}, t_2) \gamma_\nu \hat{A}_I^\nu(\boldsymbol{y}, t_2) \hat{\psi}_I(\boldsymbol{y}, t_2)] | \Psi, \boldsymbol{p} \rangle \end{aligned} \tag{29}$$

There are many terms arising from Eq. (29), most of them are originally from interactions between electrons inside of $|\Psi\rangle$, while leaving P as unaffected, that is,

$$\langle \Psi' | \int d^3\boldsymbol{x} d^3\boldsymbol{y} [e^2 \hat{\bar{\psi}}_I(\boldsymbol{x}, t_1) \gamma_\mu \hat{A}_I^\mu(\boldsymbol{x}, t_1) \hat{\psi}_I(\boldsymbol{x}, t_1) \hat{\bar{\psi}}_I(\boldsymbol{y}, t_2) \gamma_\nu \hat{A}_I^\nu(\boldsymbol{y}, t_2) \hat{\psi}_I(\boldsymbol{y}, t_2)] | \Psi \rangle \langle \boldsymbol{p}' | \boldsymbol{p} \rangle \tag{30}$$

which do not contribute to the evolution of P. Thus, we can see that the underlying mechanism would not be different with the scattering between two electrons described by QED theory. The interaction between the states $|\Psi\rangle$ and $|\boldsymbol{p}\rangle$ is mediated by the virtual photons propagating between them. Now it is clear that, in the AB effect, it does not matter whether the EMF is zero or not in the region where P can enter. The underlying mechanism is the interactions between P and $|\Psi\rangle$ while such interactions are mediated by virtual photons.

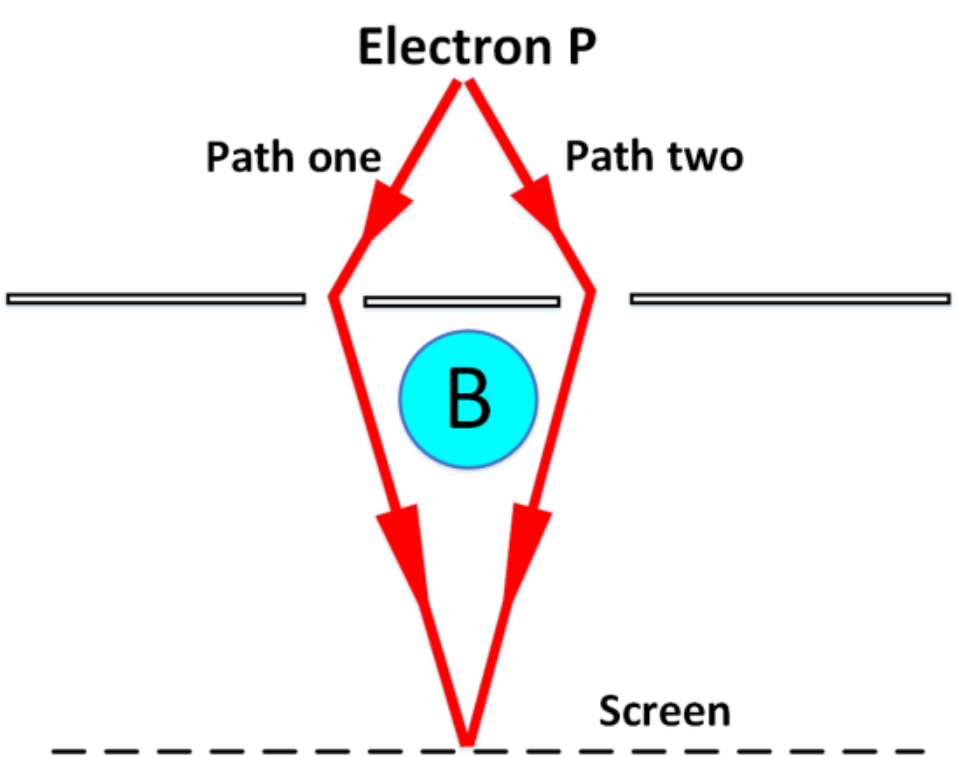

Fig. 2. Sketch of a double-slit experiment in which the Aharonov-Bohm effect can be observed. Before the magnetic field is turned on in the solenoid, the electron P with wave-vector $\boldsymbol{k}$ traveling along two different paths is in a super-positioned state as $\psi_{\boldsymbol{k}}(\boldsymbol{x}) = \exp(i\boldsymbol{k}\cdot\boldsymbol{x}_1) + \exp(i\boldsymbol{k}\cdot\boldsymbol{x}_2)$, the phase difference is $\Delta\theta = \boldsymbol{k}\cdot(\boldsymbol{x}_1 - \boldsymbol{x}_2)$. After turning on the magnetic field, the state of the electron P becomes $\psi'(\boldsymbol{x}) = \exp[i(\boldsymbol{k}\cdot\boldsymbol{x}_1 - \int_0^{x_1} e\boldsymbol{A}(\boldsymbol{x}')\cdot d\boldsymbol{x}')] + \exp[i(\boldsymbol{k}\cdot\boldsymbol{x}_2 - \int_0^{x_2} e\boldsymbol{A}(\boldsymbol{x}')\cdot d\boldsymbol{x}')]$, the phase difference will acquire a shifted value $\Delta\theta' = e\Phi_m$ in which $\Phi_m$ is the total magnetic flux in the solenoid. This phase shift can be observed as a shifted interference pattern on the screen.

In summary, since the basic building blocks of nature are mathematically constructed as relativistic fields in QFT framework we believe that the macroscopic phenomena of classical and OPQM theories originally arise from the collective effects among these fundamental quantum fields. With this belief, we reviewed physical theories from the microscopic world to the macroscopic world, that is from Eq. (1) to Eq. (5) then from Eq. (A4) to Eq. (A1), i.e., from QED to OPQM then from OPQM to classical physics. Now it has been clearly shown that the physical quantities such as, the classical potential $A^\mu(\boldsymbol{x})$ and EMF introduced from Eq. (25), were developed step by step. Since we demonstrated that the OPQM is just an approximated model which arises from quantum field theory, therefore all the physical phenomena predicted by OPQM can find their counterpart explanations in quantum field theory, including the AB effect. Actually, the non-local feature in the AB effect can be interpreted as the manifestation of virtual photons propagating between electrons in the framework of QED. Finally, we note that the introduction of EMF from Eq. (25) is not necessary in both QED and OPQM theories, however, it becomes essential only in classical physics as shown in Eq. (A3). The EMF is directly linked with velocity and acceleration of macroscopic objects, which can be easily measured.

**Acknowledgment**

We thank Prof. H. Dieter Zeh, Dr. Sivasankara Rao Yemineni and Dr. Tao Liu for helpful suggestions; we also thank Shaoxiang Chen for drawing the figures and revising the paper.

## Appendix A

In this appendix, we are further stressing on the nature of classical electromagnetic fields and the relationships between classical fields with quantum fields. In classical theory, the Hamiltonian of a charged particle in the presence of classical potential $A_\mu(t,\boldsymbol{x})$ can be given as

$$H=\frac{1}{2m}[\boldsymbol{p}-q\boldsymbol{A}(t,\boldsymbol{x})]^2+q\phi(t,\boldsymbol{x}) \tag{A1}$$

This gives the Lorentz equation of motion as

$$m\frac{d^2\boldsymbol{x}}{d^2t}=q[-\vec{\nabla}\phi-\frac{\partial\boldsymbol{A}}{\partial t}+\vec{\nabla}(\boldsymbol{v}\cdot\boldsymbol{A})-(\boldsymbol{v}\cdot\vec{\nabla})\boldsymbol{A}] \tag{A2}$$

This equation can be written in a more elegant form by introducing EMF defined in Eq. (25) as

$$m\frac{d^2\boldsymbol{x}}{d^2t}=q\boldsymbol{E}(t,\boldsymbol{x})+q\boldsymbol{v}\times\boldsymbol{B}(t,\boldsymbol{x}) \tag{A3}$$

Therefore, the necessity for the introduction of EMF through Eq. (25) which also induces people to believe that EMF corresponds to real entities of nature lies inside of Eq. (A3). As we can see, the fields $\boldsymbol{E}(t,\boldsymbol{x})$ and $\boldsymbol{B}(t,\boldsymbol{x})$ can be uniquely identified through Eq. (A3) with the information of the acceleration and velocity of the charged particle. This makes physicists to believe that the functions $\boldsymbol{E}(t,\boldsymbol{x})$ and $\boldsymbol{B}(t,\boldsymbol{x})$ corresponds to some real entities of nature since they seem to be uniquely valued at every space-time point. However, this follows the belief that the acceleration and velocity of the charged particle corresponds to some real quantities and can be uniquely valued at every space-time point along with the trajectory. In order to uniquely identify one quantity we need to identify another quantity uniquely since these quantities are bonded together in one equation. Measuring each quantity $\boldsymbol{E}(t,\boldsymbol{x})$ and $\boldsymbol{B}(t,\boldsymbol{x})$ precisely at every space-time point requests us to treat the particle as a single space-time point with zero size. This condition is too unrealistic to be satisfied in classical physics;

it demands more internal structures of the macroscopic particle. However, we know that for microscopic particles, the velocity along a trajectory is not well defined due to the uncertainty principle. This makes Eq. (A3) as well as the bond break down in the micro-world. Therefore, we replace $\boldsymbol{p}$ with $-i\vec{\nabla}$ in Eq. (A1) and obtain the Hamiltonian of an electron in the non-relativistic limit as

$$H = \frac{1}{2m}[-i\vec{\nabla} + e\boldsymbol{A}(t,\boldsymbol{x})]^2 - e\phi(t,\boldsymbol{x}) \quad \text{(A4)}$$

In radiation gauge, we get the wave-function of the electron as $|\psi\rangle = \exp(-i\int_0^{\boldsymbol{x}} e\boldsymbol{A}(\boldsymbol{x}')\cdot d\boldsymbol{x}')|\boldsymbol{p}\rangle$ in case of a non-time-varying vector potential. For the AB effect depicted in Fig. 2, the electron traveling enclosing a circle will pick up a phase shift that can be measured which is $\int_S e\boldsymbol{B}(\boldsymbol{x})\cdot d\boldsymbol{S} = e\Phi_m$, where $\Phi_m$ is the total magnetic flux through the closed surface. However, it reminds that two wave-functions with the phase difference of $2n\pi$ ($n = \pm 1, \pm 2 \cdots$) still cannot be distinguished. Therefore, neither classical physics nor quantum physics can uniquely quantify $\boldsymbol{E}(t,\boldsymbol{x})$ and $\boldsymbol{B}(t,\boldsymbol{x})$ at each space-time point precisely. This comes as expected actually since we already argued that the classical potential $A_\mu(t,\boldsymbol{x})$ together with classical EMF defined by Eq. (25) are emergent properties and arise from QED process. In Eq. (A3), EMF is introduced to approximate the interactions between charged particles governed by quantum physics.

At this stage, we argue that different physical quantities along with different theories arise at different spatial-temporal scales. The most fundamental nature law at the deepest level may be unique; however, the ignorance of the detailed structures at smaller scales permits physicists to create theories, which are approximately effective at larger scales. Moreover, physicists create mathematical equations which predict the evolutions of nature, and the mathematical form of the physical quantities on both sides of the equation needs to be constructed consistently in order to fit the equation form. As we can see from our derivations of the classical potentials given in the Eq. (21), this mathematical form of the classical four-potential is constructed to fit the framework of OPQM in order to give the same predictions with QED theory in low energy limit. Hence, in the view of physical laws from micro-world to macro-world, the emergence of $A_\mu(t,\boldsymbol{x})$ follows the mathematical construction of the framework of OPQM, while the EMF follows the Lorentz force equation, or in other words we can say that these quantities are bonded with the framework of OPQM and framework of Lorentz force equation, respectively. Therefore, these quantities cannot be divorced from their frameworks, and the nature does not specify what these quantities are without referring to what roles that they are playing in the frameworks. To be more specific, let us just simply multiply by 2 on both sides of Eq. (A3) and rescale the quantities $m' = 2m$, $\boldsymbol{E}'(t,\boldsymbol{x}) = 2\boldsymbol{E}(t,\boldsymbol{x})$ and $\boldsymbol{B}'(t,\boldsymbol{x}) = 2\boldsymbol{B}(t,\boldsymbol{x})$ such that the equivalence relation with the rescaled quantities still holds [that is, if we revalue the mass of every macroscopic object in our universe, the EMF has to be revalued accordingly]. In this way, the new equation with the rescaled quantities works just as good as the old one, but nature does not tell us which one we should use, and which quantity that is, $\boldsymbol{E}'(t,\boldsymbol{x})$ or $\boldsymbol{E}(t,\boldsymbol{x})$ should be defined as the real physical entity of nature. Therefore, in this case we can safely speak that, at this macroscopic scale, only the equivalence relation in Eq. (A3) is the concrete thing that we should stick with. Any transformation of Eq. (A3) with new defined quantities must mathematically maintain such equivalent relation in order to give the same physical measurement predictions. This is also what happens from Eq. (A2) to Eq. (A3). Alternatively, we can rescale the strength of EMF and the charge $q$ instead of $m$ in Eq. (A3), and then the above argument also applies. Moreover, quantities defined at one scale may break down at another scale, such as color or the temperature of an object which cannot be well defined at the microscopic scale since they are originally from something else that are more fundamental. The similar argument applies to the classical potential $A_\mu(t,\boldsymbol{x})$ and EMF which both arise originally from low energy QED physics. In addition, EMF role is insignificant in the mathematical constructions of OPQM and QED frameworks, QED and OPQM are complete theories even without the EMF as we can see in Eq. (1) and Eq. (5). Next we are going to provide another evidence which reaffirm our statement.

Suppose that there exists a static classical field $\boldsymbol{E}(\boldsymbol{x}) \neq 0$ and $\boldsymbol{B}(t,\boldsymbol{x}) = 0$ somewhere in the "source" free region, for simplicity, we assume that the polarization of $\boldsymbol{E}(\boldsymbol{x})$ is in the $z$ direction in the reference frame $(t,\boldsymbol{x})$, i.e., $E_x(\boldsymbol{x}) = E_y(\boldsymbol{x}) = 0$. Therefore, the static field $E_z(\boldsymbol{x})$ can be given as

$$E_z(\boldsymbol{x}) = i\int \frac{d\omega d^3\boldsymbol{k}}{(2\pi)^3}[\tilde{E}_z(\omega,\boldsymbol{k})e^{-i(\omega t-\boldsymbol{k}\cdot\boldsymbol{x})} - \tilde{E}_z^\dagger(\omega,\boldsymbol{k})e^{i(\omega t-\boldsymbol{k}\cdot\boldsymbol{x})}] \tag{A5}$$

in order to get a non-time-varying function $E_z(\boldsymbol{x})$, we require $\tilde{E}_z(\omega,\boldsymbol{k}) = \delta(\omega)\tilde{f}(\boldsymbol{k})$ in which $\delta(\omega)$ is the Dirac-Delta function and $\tilde{f}(\boldsymbol{k})$ is a function of wave-vector $\boldsymbol{k}$. Next, we perform a Lorentz boost with velocity $v$ in $z$ direction and obtain a new field $E'_z(t',\boldsymbol{x}')$ written in $(t',\boldsymbol{x}')$ frame with relation

$$E'_z(t',\boldsymbol{x}') = E_z(t,\boldsymbol{x}) \tag{A6}$$

The modes expansion of $E'_z(t',\boldsymbol{x}')$ can be given as

$$E'_z(t',\boldsymbol{x}') = i\int \frac{d\omega' d^3\boldsymbol{k}'}{(2\pi)^3}[\tilde{E}'_z(\omega',\boldsymbol{k}')e^{-i(\omega't'-\boldsymbol{k}'\cdot\boldsymbol{x}')} - \tilde{E}'^{\dagger}_z(\omega',\boldsymbol{k}')e^{i(\omega't'-\boldsymbol{k}'\cdot\boldsymbol{x}')}] \tag{A7}$$

In the new reference frame we have $\begin{Bmatrix} t' = \gamma(t+vz) \\ z' = \gamma(z+vt) \end{Bmatrix}$ and $\begin{Bmatrix} \omega' = \gamma(\omega+vk_z) \\ k'_z = \gamma(k_z+v\omega) \end{Bmatrix}$ with $\gamma = 1/\sqrt{1-v^2}$. Therefore, we can obtain the relation $\tilde{E}_z(\omega,\boldsymbol{k}) = \tilde{E}'_z(\omega',\boldsymbol{k}')$ as a result of Eq. (A6) and $e^{i(\omega't'-\boldsymbol{k}'\cdot\boldsymbol{x}')} = e^{i(\omega t-\boldsymbol{k}\cdot\boldsymbol{x})}$, i.e., the Fourier components of field $E_z(\boldsymbol{x})$ does not change in the new reference frame. This result is also what we expect in quantum field theory: the probability corresponding to measurement outcomes must be a Lorentz invariant. Plug relation $\tilde{E}'_z(\omega',\boldsymbol{k}') = \delta(\omega)\tilde{f}(\boldsymbol{k})$ into Eq. (A7), after integrating over frequency $\omega'$ we find a nonzero value at $\omega' = \gamma v k_z$. Therefore, we have brought a non-time-varying field $E_z(\boldsymbol{x})$ into a time-varying field $E'_z(t',\boldsymbol{x}')$ by a Lorentz boost. Furthermore, we noticed that the phase velocity of the modes $e^{-i(\omega't'-\boldsymbol{k}'\cdot\boldsymbol{x}')}$ in $E'_z(t',\boldsymbol{x}')$ can be given as $u' = \frac{\gamma v k_z}{|\boldsymbol{k}'|}$ measured in the reference frame $(t',\boldsymbol{x}')$, therefore, the modes which comprise the field $E'_z(t',\boldsymbol{x}')$ are propagating at speed $u' \le v$ which is slower than light. Thus, the modes in $E'_z(t',\boldsymbol{x}')$ cannot be photons. In fact, neither $E'_z(t',\boldsymbol{x}')$ nor $E_z(\boldsymbol{x})$ can be quantized. Or, at least we can say, the "quantization" of these classical fields and classical potentials do not give rise to particles behaving like photons. The reason is, as shown in our main text of this article, the EMF are emergent quantities and are not directly linked with some elementary particles, i.e., photons in this scenario.

For the FEMF introduced following Eq. (2) and Eq. (25), we have

$$\begin{aligned}
\boldsymbol{E}(t,\boldsymbol{x}) &= i\int \frac{d^3\boldsymbol{k}}{(2\pi)^3}\sqrt{\frac{\omega_{\boldsymbol{k}}}{2}}\sum_{\lambda=1,2}[\tilde{E}^{\lambda}_{\boldsymbol{k}}\boldsymbol{v}^{\lambda}e^{-i(\omega_{\boldsymbol{k}}t-\boldsymbol{k}\cdot\boldsymbol{x})} - \tilde{E}^{\lambda\dagger}_{\boldsymbol{k}}\boldsymbol{v}^{\lambda}e^{i(\omega_{\boldsymbol{k}}t-\boldsymbol{k}\cdot\boldsymbol{x})}] \\
\boldsymbol{B}(t,\boldsymbol{x}) &= i\int \frac{d^3\boldsymbol{k}}{(2\pi)^3}\frac{1}{\sqrt{2\omega_{\boldsymbol{k}}}}\sum_{\lambda=1,2}[\tilde{B}^{\lambda}_{\boldsymbol{k}}\boldsymbol{k}\times\boldsymbol{v}^{\lambda}e^{-i(\omega_{\boldsymbol{k}}t-\boldsymbol{k}\cdot\boldsymbol{x})} - \tilde{B}^{\lambda\dagger}_{\boldsymbol{k}}\boldsymbol{k}\times\boldsymbol{v}^{\lambda}e^{i(\omega_{\boldsymbol{k}}t-\boldsymbol{k}\cdot\boldsymbol{x})}]
\end{aligned} \tag{A8}$$

Note that these quantities above are totally different from the EMF quantities in Eq. (A3). In Eq. (A8), the FEMF is defined from a Lorentz vector in the quantized form as Eq. (2).

Therefore, the FEMF are just two different mathematical constructions that are made up of photons, and they, as a result, form a real Lorentz tensor. In addition, the FEMF also plays an essential role in the mathematical formulation of QED theory. However, there is no counterpart of EMF in quantum physics and the EMF emerges in macro-world due to the collective effects from micro-world. Now we see that the FEMF and EMF possess totally different physical meanings, the FEMF is made up of photons which are elementary particles of nature while EMF only serves as a mathematical tool in classical physics. The difference in physical meanings between FEMF and EMF originates from the differences between $\hat{A}_\mu$ in Eq. (1) and $A_\mu(\boldsymbol{x})$ in Eq. (5) which is a derived quantity; indeed it would be less confusing if physicists historically have denoted the EMF and FEMF using two different symbols since they hold different physical meanings. Moreover, the mathematical structure of $\hat{A}_\mu$ and FEMF are fixed, while $A_\mu(\boldsymbol{x})$ and EMF appear to be completely arbitrary in the real world since they depend on the sources. Finally, we should be more careful over the differences of their mathematical structures rather than what historically people have symbolized them since only their mathematical structures tell us what these quantities really are.